\pdfoutput=1
\documentclass[10pt,twocolumn]{article}
\usepackage[a4paper,margin=1.8cm]{geometry}
\usepackage{amsmath, amssymb}
\usepackage{graphicx}
\usepackage{cuted}
\usepackage{hyperref}
\usepackage{adjustbox}  
\usepackage{titlesec}
\usepackage{lmodern}
\usepackage{microtype}
\usepackage{booktabs} 
\usepackage{times}
\usepackage{subcaption}
\usepackage{calc}
\graphicspath{{figures/}{plots_out/}}

\title{RL-Exec: Impact-Aware Reinforcement Learning for Opportunistic Optimal Liquidation,
Outperforms TWAP and a Book-Liquidity VWAP on BTC-USD Replays}
\author{Enzo Duflot, Stanislas Robineau}
\date{October 2025}

\begin{document}

\maketitle

\section{Abstract}

We study opportunistic optimal liquidation over fixed deadlines on BTC-USD limit-order books (LOB). We present RL-Exec, a PPO agent trained on historical replays augmented with endogenous transient impact (resilience), partial fills, maker/taker fees and latency. The policy observes depth-20 LOB features plus microstructure indicators and acts under a sell-only inventory constraint to reach a residual target. \\

Evaluation follows a time split (train: Jan-2020; test: Feb-2020) and a per-day protocol: for each test day we run 10 independent start times within the day and aggregate to a single daily score, avoiding pseudo-replication. We compare the agent to (i) TWAP and (ii) a VWAP-like baseline that allocates using order-book liquidity on the opposite side (top-20 levels), both executed on exactly the same timestamps and costs as the agent. Statistical inference uses a one-sided Wilcoxon signed-rank test on daily RL--baseline differences with Benjamini--Hochberg FDR correction and bootstrap confidence intervals. \\

On the Feb-2020 test set, RL-Exec significantly outperforms both baselines and the gap increases with the execution horizon: \\

\begin{table}[h!]
\raggedright
\setlength{\tabcolsep}{4pt} 
\small
\caption{Estimated effects by duration with adjusted $p$-values and bootstrap 95\% CIs.}
\begin{adjustbox}{max width=\linewidth}
\begin{tabular}{@{}lccc@{}}
\toprule
\textbf{Duration (s)} & \textbf{Effect (bps)} & \textbf{$p_{\mathrm{adj}}$} & \textbf{95\% CI (bootstrap)} \\ 
\midrule
1,800  & +2.25 -- 2.68   & $<0.01$                & positive \\
3,600  & +7.59 -- 7.70   & $\approx 3$--$5\times10^{-6}$ & positive \\
7,200  & +22.96 -- 23.02 & $\approx 1\times10^{-8}$     & positive \\
\bottomrule
\end{tabular}
\end{adjustbox}
\end{table}

\textit{Code:} \href{https://github.com/Giafferri/RL-Exec}{github.com/Giafferri/RL-Exec}.

\section{Introduction}

Execution desks still default to simple schedules (TWAP/VWAP) because they're robust and capacity-friendly, yet they ignore microstructure state and typically realize negative PnL once spreads and temporary impact are accounted for. Learning-based approaches promise opportunistic timing, but many studies (i) evaluate in exogenous replays that ignore the agent's footprint, (ii) use weak or mismatched baselines, or (iii) lack per-day statistical testing, making significance claims fragile. \\

We address these gaps with RL-Exec, an impact-aware RL framework for sell-only liquidation on BTC-USD LOBs. The environment replays historical books while injecting transient impact with resilience, partial fills, maker/taker fees, and latency so that actions affect execution prices and fills. Policies observe depth-20 features plus standard microstructure indicators and are trained with PPO. \\

Evaluation uses a strict time split (Jan $\to$ train, Feb $\to$ test) and a per-day protocol with multiple intra-day starts aggregated to a single daily score, enabling paired, one-sided Wilcoxon tests with FDR control and bootstrap CIs. Baselines are run on exactly the same timestamps and cost model: TWAP and a VWAP-like, liquidity-weighted schedule (book-based approximation given the replay setting). \\

\textbf{Contributions.} Impact-aware replay for execution. Fast, reproducible LOB environment with endogenous transient impact, partial fills, and exchange frictions; usable for both RL and baseline schedules on identical windows. \\

\textbf{Rigorous per-day inference.} Deterministic daily protocol with multiple starts/day, Wilcoxon signed-rank tests (one-sided), Benjamini--Hochberg correction, and bootstrap confidence intervals. \\

\textbf{Opportunistic liquidation policy.} A PPO agent that beats TWAP and the liquidity-weighted VWAP proxy on out-of-sample days; the advantage grows with horizon (details in Results). \\

\textbf{Reproducibility.} Public commands produce per-episode/day CSVs and Markdown summaries. \\

Results cover a single asset (BTC-USD), sell-only liquidation, and one out-of-sample month. The VWAP comparator is an order-book-liquidity proxy (not true traded-volume VWAP), and the impact process is parametric. Extending to buys, multi-venue routing, alternative impact calibrations, and policy stress-tests under regime shifts are left for follow-up work.

\section{Related Work}

\textbf{Optimal execution under impact.} Classical models frame execution as a cost--risk trade-off with market impact. Almgren \& Chriss (2001) provide risk-aware optimal schedules under linear temporary/permanent impact. Resilient LOB models (Obizhaeva \& Wang, 2013) and propagator kernels (Gatheral \& Schied) formalize transient impact and no-manipulation conditions, aligning with empirical concavity (e.g., square-root). These strands motivate simulators that include both immediate slippage and post-trade resilience---assumptions we adopt in RL-Exec. \\

\textbf{Industry baselines.} TWAP/VWAP remain ubiquitous for capacity and simplicity but are state-agnostic and typically realize negative expected PnL once spreads/impact are counted. More adaptive heuristics include POV (fixed or signal-aware) and Almgren--Chriss schedules calibrated to venue costs and volatility. In our replay setting we compare to TWAP and a VWAP-like liquidity proxy (book-weighted allocation) executed on exactly the same timestamps and fee/impact model as the agent. \\

\textbf{RL for execution.} Early work (Nevmyvaka, Feng \& Kearns, 2006) showed that order-book features (spread, imbalance, short-horizon markout) enable RL to beat static schedules on historical equities. Subsequent deep-RL studies (e.g., actor--critic/PPO variants) often report gains over TWAP/VWAP, but many evaluate in exogenous replays without agent footprint, use weak comparators, or lack per-day statistical testing---risking optimistic conclusions. RL-Exec addresses these gaps with impact-aware replay, strong book-based baselines on identical windows, and paired, one-sided Wilcoxon tests with FDR control and bootstrap CIs. \\

\textbf{Execution environments.} Agent-based simulators (e.g., ABIDES) endogenize impact via interacting agents but can be heavy to calibrate. Hybrid replay-with-impact environments strike a pragmatic balance by injecting transient impact and replenishment on top of real event streams, preserving empirical flow while making the agent pay for footprint---the design we use. \\

\textbf{Crypto microstructure \& data.} Crypto markets are 24/7, fee-tiered, and fragmented, with replenishment and maker rebates that materially affect execution. Our experiments use BTC-USD LOB data with depth-20 features; the dataset and preliminary analysis were provided by Jonathan Sadighian (SESAMm) from Extending Deep Reinforcement Learning Frameworks in Cryptocurrency Market Making, which informed our feature engineering and evaluation protocol. 

Comparable sets of LOB-based indicators have been successfully employed in reinforcement learning frameworks such as \textit{DeepLOB}~\cite{zhang2018deeplob} and \textit{Deep Reinforcement Learning for Market Making}~\cite{zhang2021deep}, demonstrating their effectiveness in capturing microstructural dynamics and improving policy learning in high-frequency trading environments.

\section{Data and Preprocessing}

We use BTC-USD limit-order-book data at 1 s resolution with 20 levels per side. Training covers January 2020 (31 days) and testing covers February 2020 (28 days). The dataset and preliminary analysis were provided by Jonathan Sadighian (SESAMm), Extending Deep Reinforcement Learning Frameworks in Cryptocurrency Market Making. From the raw event stream we construct 1 s snapshots containing the top-20 bid/ask ladders, compute mid-price and spread, forward-fill missing levels with zero size, and enforce strictly increasing timestamp\_ns. Each day is stored as a Feather file (\texttt{data/replay\_merged/YYYYMMDD.feather}) for fast replay. \\

Observations follow the depth20+ind specification: prices are normalized by mid, sizes are expressed in BTC, and we append a fixed set of microstructure indicators (spread, imbalance, short-horizon returns/volatility, microprice-style terms). The split is strictly temporal (no shuffling). We fit VecNormalize on January observations (reward normalization off), save its statistics, and reload the same transform for all February evaluations to avoid leakage. \\

Basic quality filters remove duplicate timestamps and rows with invalid best bid/ask or non-positive spread; obviously nonsensical prices/sizes are clipped. Evaluation is per-day: for each test day we run 10 distinct intra-day start times and a fixed horizon (1,800 / 3,600 / 7,200 s). The RL agent and baselines are executed on the exact same timestamps and cost model, and we aggregate the 10 runs to a single daily score to avoid pseudo-replication. \\

Baselines include TWAP and a VWAP-like proxy that allocates in proportion to opposite-side order-book depth over the top-20 levels (a book-based approximation appropriate for replay; it is not true traded-volume VWAP). All training, evaluation, and statistical tests are scriptable and reproducible. Known limitations of this setup include parametric transient impact with resilience (no other agents), a single asset, sell-only liquidation, and the book-based VWAP proxy.

\section{Execution Environment}

We evaluate policies in a fast replay-with-impact environment that streams historical BTC--USD order books while making the agent ``pay'' for its footprint. Each episode is a fixed intraday window defined by a start index and a horizon (1,800 / 3,600 / 7,200 s). At every step the environment pulls the depth-20 book at the current nanosecond timestamp, applies the agent's order, settles fills with fees and impact, advances time by one snapshot, and carries over the updated inventory and cash. \\

Observations follow the depth20+ind specification: top-20 bid/ask ladders (prices normalized by mid, sizes in BTC) augmented with standard microstructure indicators (e.g., spread, imbalance, short-horizon returns/volatility, microprice-style terms), plus remaining-time and remaining-inventory fractions. We fit VecNormalize on January data and reuse the frozen statistics at test time. \\

Actions implement sell-only liquidation. The policy either holds or submits a sell of bounded rate; the per-step notional is capped by a configurable trade fraction of remaining inventory. To focus the study on liquidation, opposite-side trades are disabled during evaluation (no buys while selling). Orders are executed against the displayed book with partial fills: the simulator consumes depth level-by-level until either the order is filled or available liquidity is exhausted. \\

Costs and impact mirror exchange frictions and transient footprint. We account for maker rebates and taker fees, as well as slippage from walking the book. Temporary price impact shifts the effective execution level and then decays with resilience: an immediate impact term proportional to executed size (concave in size and scaled by contemporaneous liquidity) is applied at execution and reverts exponentially over subsequent steps. This yields realistic short-horizon mark-outs and prevents cost underestimation in aggressive bursts. A small, fixed latency is included so that decisions affect prices one tick later, which avoids optimistic fill timing. \\

Rewards align with implementation shortfall under a hard deadline. Per-step rewards track execution PnL relative to the mid-price reference, with a terminal mark-to-market of any residual position at the episode end. A mild penalty on remaining inventory discourages missing the target. Episodes terminate at the horizon; there are no resets within a day. \\

Baselines are run on the exact same timestamps and cost model as the agent to enable paired testing. TWAP splits the target evenly across steps; the VWAP-like proxy allocates proportionally to opposite-side book depth over the top-20 levels (a book-based approximation appropriate in replay where true traded volume is unavailable). \\

Determinism is ensured by fixing seeds, freezing normalization stats, and using a per-day protocol that aggregates multiple intra-day starts to a single daily score for significance testing. Limitations of this setup include a single asset, a parametric impact/resilience process rather than interacting agents, and the VWAP proxy's reliance on displayed depth rather than realized volume.

\section{Baselines}

We compare RL-Exec to two schedule-based baselines implemented inside the same replay-with-impact simulator and executed on exactly the same timestamps, fee model, and impact/resilience process as the agent. This ensures paired, like-for-like evaluation. \\

\textbf{TWAP (time-weighted average price).} \\

Given a fixed horizon of \(N\) steps and a sell-only target of \((1-\text{target})\times \texttt{initial\_btc}\), TWAP submits an equal BTC amount each step: \(q_t = Q/N\). Orders trade against the displayed bid ladder with partial fills; if available depth is insufficient, the unfilled remainder is not rolled over (i.e., the plan is myopic), which mirrors many simple schedule engines. Costs include taker/maker fees, slippage from walking the book, and transient impact with resilience. Mark-to-market at the end uses the same pricing as RL-Exec. \\

\textbf{VWAP-like (book-weighted allocation).} \\

True traded volume is unavailable in replay, so we use a liquidity-weighted proxy: at step \(t\), we compute weights from the opposite-side liquidity in the top-\(L\) bid levels (default \(L=20\)):
\[
w_t \propto \sum_{\ell=1}^{L} \mathrm{size}^{\mathrm{BTC}}_{t,\ell}
\]
(falling back to \(\mathrm{notional}^{\mathrm{USD}}/\mathrm{mid}\) if needed). The schedule allocates \(Q\) proportionally, \(q_t = Q \cdot \frac{w_t}{\sum_{\tau} w_{\tau}}\). Orders execute identically to TWAP (partial fills, same costs/impact). This baseline adapts to displayed depth without look-ahead and is a reasonable approximation of VWAP in a replay setting. \\

\textbf{Fairness \& protocol.} \\

Both baselines (i) use the same episode window as RL-Exec (same start index, duration, and per-day seeds), (ii) face identical frictions (fees, latency, slippage, transient impact with resilience), and (iii) are evaluated with opposite-side trades disabled during liquidation to match the agent's sell-only constraint. Results are recorded per episode and aggregated per day (mean over 10 intra-day starts) before statistical testing. \\

\paragraph{Scope of baselines.}
We do not claim to outperform calibrated Almgren--Chriss (AC) or Participation-of-Volume (POV) schedules; that comparison requires venue-calibrated risk/impact parameters and is left to future work. Our aim here is narrower: to show that an impact-aware opportunistic policy can reliably beat simple time/liquidity schedules in a like-for-like replay.

\section{Order Book Indicators}

In this work, we extract a comprehensive set of limit order book (LOB) indicators to represent both the instantaneous market state and its short-term dynamics. These indicators serve as inputs to the reinforcement learning (RL) agent, allowing it to learn profitable trading policies based on microstructural signals. \\

\subsection{Micro-price}

The \textit{micro-price} provides a refined estimate of the fair market value by weighting the best bid and ask prices according to their respective order sizes: \\
\[
p_{\text{micro}} = \frac{p_{\text{ask}}\, Q_{\text{bid}} + p_{\text{bid}}\, Q_{\text{ask}}}{Q_{\text{bid}} + Q_{\text{ask}}}
\]
where \( p_{\text{bid}} \) and \( p_{\text{ask}} \) are the best bid and ask prices, and \( Q_{\text{bid}} \), \( Q_{\text{ask}} \) are their corresponding quantities. \\
A micro-price above the midpoint suggests buying pressure, while a lower value indicates selling pressure. \\

This indicator has been extensively used in high-frequency trading models such as \textit{DeepLOB}~\cite{zhang2018deeplob}. \\

\subsection{Order Book Imbalance}

The \textit{top-of-book imbalance} quantifies the asymmetry of liquidity between buy and sell sides:
\[
I = \frac{Q_{\text{bid}} - Q_{\text{ask}}}{Q_{\text{bid}} + Q_{\text{ask}}}
\]
with \( I > 0 \) indicating buying pressure and \( I < 0 \) selling pressure. \\
The multi-level imbalance extends this measure to several levels of the book:
\[
I_n = \frac{\sum_{i=1}^{n} Q_{\text{bid}}^{(i)} - \sum_{i=1}^{n} Q_{\text{ask}}^{(i)}}{\sum_{i=1}^{n} Q_{\text{bid}}^{(i)} + \sum_{i=1}^{n} Q_{\text{ask}}^{(i)}}
\]
These indicators are strong predictors of short-term price movements~\cite{cartea2020algorithmic, sirignano2019deep}. \\

\subsection{Spread and Liquidity Metrics}

The \textit{normalized spread} measures market tightness and transaction cost: \\
\[
S_{\text{norm}} = \frac{p_{\text{ask}} - p_{\text{bid}}}{p_{\text{mid}}} \times 100
\]
where \( p_{\text{mid}} = \frac{p_{\text{ask}} + p_{\text{bid}}}{2} \) is the midpoint price. \\
The \textit{bid depth} and \textit{ask depth} are defined as: \\
\[
D_{\text{bid}} = \sum_i Q_{\text{bid}}^{(i)}\, p_{\text{bid}}^{(i)}, \qquad D_{\text{ask}} = \sum_i Q_{\text{ask}}^{(i)}\, p_{\text{ask}}^{(i)}.
\]
These metrics provide information about liquidity concentration and help the agent adjust its order aggressiveness. \\

\subsection{Volume Adjusted Mid-Price (VAMP)} 

The \textit{VAMP} represents a volume-weighted equilibrium price, reducing noise compared to the raw midpoint:
\[
\mathrm{VAMP} = \frac{\frac{\sum_i Q_{\text{ask}}^{(i)} p_{\text{ask}}^{(i)}}{\sum_i Q_{\text{ask}}^{(i)}} + \frac{\sum_i Q_{\text{bid}}^{(i)} p_{\text{bid}}^{(i)}}{\sum_i Q_{\text{bid}}^{(i)}}}{2}.
\]
This indicator captures the balance between traded volume and price, providing a stable estimate of fair market value. 

\subsection{Order Flow Imbalance (OFI)} 

The \textit{Order Flow Imbalance} measures changes in liquidity between two consecutive timestamps:
\[
\mathrm{OFI} = \frac{\Delta Q_{\text{bid}} - \Delta Q_{\text{ask}}}{\Delta Q_{\text{bid}} + \Delta Q_{\text{ask}}}.
\]
A positive OFI indicates net buying pressure, while a negative value suggests net selling pressure. \\

This feature reflects the dynamic flow of orders and is crucial for detecting short-term market shifts~\cite{cont2011price, zhang2021deep}. 

\subsection{Book Pressure Index (BPI)} 

The \textit{Book Pressure Index} (BPI) measures the relative strength of bid-side and ask-side liquidity, weighted by their distance to the midprice: \\
\[
\mathrm{BPI} = \frac{\sum_i \dfrac{Q_{\text{bid}}^{(i)}}{|d_{\text{bid}}^{(i)}|}}{\sum_i \dfrac{Q_{\text{ask}}^{(i)}}{|d_{\text{ask}}^{(i)}|}}
\]
where \( d_{\text{bid}}^{(i)} \) and \( d_{\text{ask}}^{(i)} \) represent the distance of each level to the midpoint. \\

A value \( \mathrm{BPI} > 1 \) suggests buying pressure, whereas \( \mathrm{BPI} < 1 \) implies selling pressure~\cite{kercheval2015modeling}. 

\subsection{Dynamic Indicators}

Temporal features such as the \(\Delta\)-midpoint and \(\Delta\)-VAMP capture short-term market momentum:
\[
\Delta p_{\text{mid}} = p_{\text{mid},t} - p_{\text{mid},t-1}, \qquad
\Delta \mathrm{VAMP} = \mathrm{VAMP}_t - \mathrm{VAMP}_{t-1}.
\]
These indicators are essential for learning reactive and adaptive trading strategies based on recent market evolution~\cite{jin2020reinforcement}.

\subsection{Correlations}

\begin{figure}[h!]
    \centering
    \includegraphics[width=0.8\linewidth]{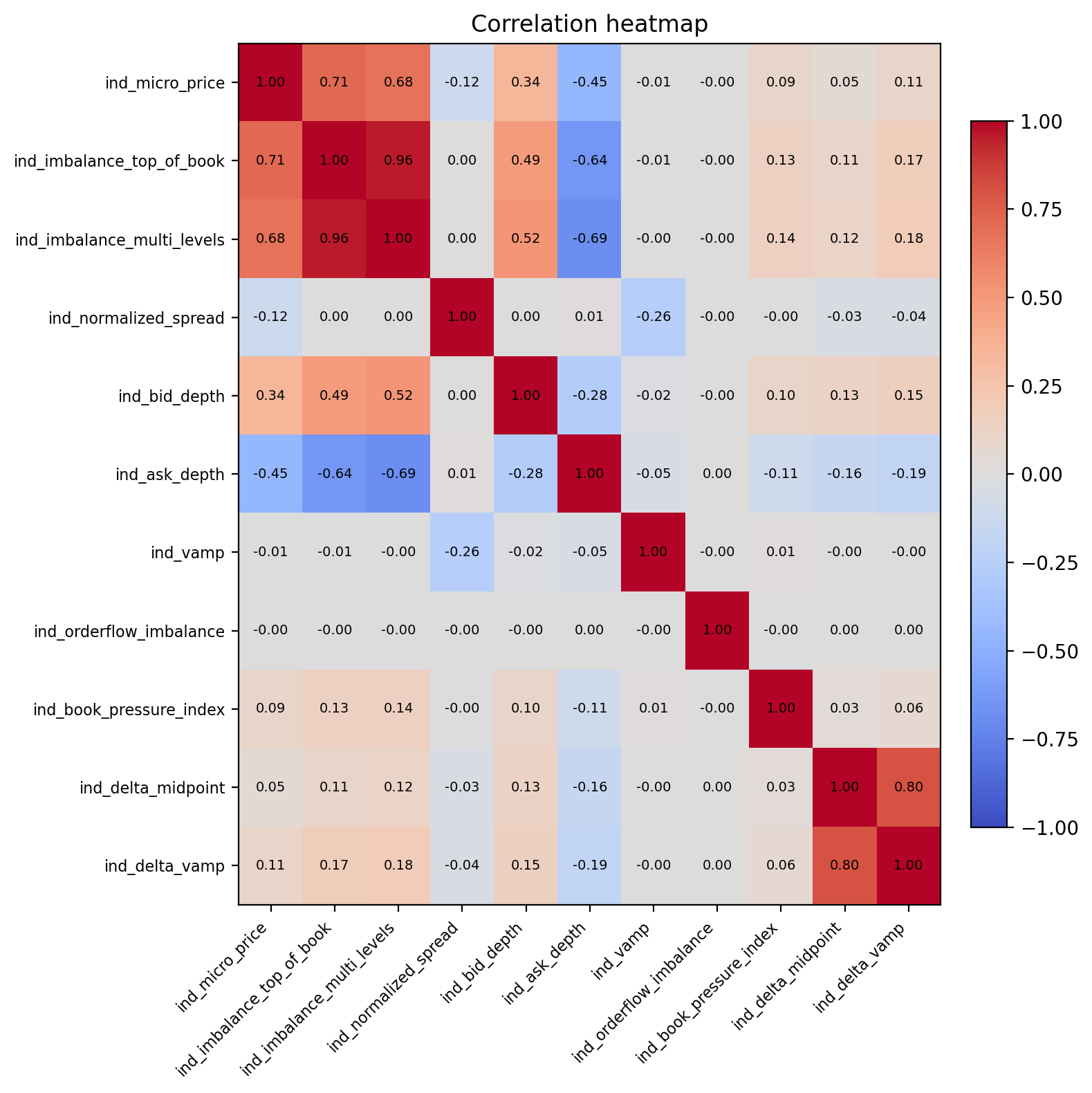}
    \caption{Indicators Correlation Heatmap.}
    \label{fig:performance}
\end{figure}

\noindent
The correlation heatmap highlights a strong cluster among imbalance-related indicators (\texttt{ind\_micro\_price}, \texttt{ind\_imbalance\_top\_of\_book}, \texttt{ind\_imbalance\_multi\_levels}), which capture similar information about book asymmetry. Bid and ask depths exhibit moderate opposite correlations with these variables, while delta features are highly interrelated. Most remaining indicators show near-zero correlations, suggesting they provide complementary and largely independent signals about market microstructure dynamics.

\section{RL Agent \& Training}

We train a PPO agent (stable-baselines3) with a two-layer MLP policy (256 and 128 units, SiLU activations) to perform sell-only liquidation. The action is a single continuous scalar interpreted as the fraction of remaining inventory to execute at the current step; it is clipped to [0,\texttt{trade\_fraction}], so negative proposals map to ``hold.'' During evaluation, opposite-side trades are disabled to enforce strict liquidation. Observations combine a depth-20 snapshot of the book (both sides) with standard microstructure indicators (spread, imbalance, short-horizon returns/volatility) and state variables (time-to-go, inventory). \\

Inputs are normalized with VecNormalize fitted on the training split and frozen for testing.
The reward mirrors implementation shortfall computed under the same fee, transient-impact, and resilience model used by the baselines, with shaping that penalizes leftover inventory and encourages hitting a residual target by the deadline. Training uses Jan-2020 replays with randomized intra-day start indices and fixed horizons; we report out-of-sample performance on Feb-2020 for 1,800/3,600/7,200-second windows. \\

PPO hyperparameters follow SB3 defaults unless noted: learning rate \(1\!\times\!10^{-4}\), clip range \(0.20\), \(\gamma=0.995\), GAE \(\lambda=0.95\), value-loss coefficient \(0.5\), entropy coefficient \(0.0\), max-grad-norm \(0.5\). We enable KL early stopping with a max-KL of \(0.05\), which halts updates once policy shifts become small---consistent with the training logs. \\

Runs use DummyVecEnv on CPU for \(\sim 10\)M steps with reward/observation normalization. We fix seeds and evaluate deterministically (greedy mean action). No validation set is used for checkpoint selection; instead, we freeze normalization statistics from training and test the final checkpoint out-of-sample. \\

Checkpoints and VecNormalize states are saved and explicitly reloaded to ensure reproducibility, and baselines are executed on exactly the same timestamps and cost/impact model as the agent.

\section{Evaluation Protocol}

We train on Jan-2020 and evaluate strictly out-of-sample on Feb-2020. For each execution horizon \(H \in \{1800, 3600, 7200\}\) seconds, we adopt a per-day design: on every test day we run the policy deterministically on ten distinct intra-day start indices, producing ten fixed-length windows. \\

TWAP and the VWAP-like liquidity proxy are executed on exactly the same timestamps and under the same fee and transient-impact/resilience settings as the agent, ensuring cost and path parity. To avoid pseudo-replication, we aggregate within day by averaging the ten pnl\% outcomes for RL and for each baseline, then form paired daily differences \(\Delta_d=\overline{\mathrm{RL}}_d-\overline{\mathrm{Baseline}}_d\); these \(\{\Delta_d\}\) constitute the units of inference.  \\

The primary endpoint is percentage PnL (pnl\%); cumulative reward is logged only for diagnostics. For each horizon we test the one-sided hypothesis that RL outperforms each baseline using the Wilcoxon signed-rank test on \(\{\Delta_d\}\), report paired t-tests as a secondary check, and compute 95\% bootstrap confidence intervals for the mean difference (10k resamples). When testing multiple baselines within a horizon, p-values are adjusted with Benjamini--Hochberg FDR; horizons are treated as separate families. \\

Reproducibility is ensured by reloading the training VecNormalize statistics, fixing random seeds, and exporting per-episode rows (day, episode id, start index, horizon, RL/TWAP/VWAP pnl\%) to CSV alongside an aggregated CSV/Markdown statistical report.

\section{Statistical Testing}

Our endpoint is percentage PnL over a fixed horizon. For each test day we run 10 independent intra-day starts, then aggregate the starts to a single daily score per method; days---not starts---are the units of inference. For each baseline \(b\) (TWAP and the VWAP-like proxy) we form paired daily differences \(\Delta_d = \overline{\mathrm{RL}}_d - \overline{b}_d\) and test the one-sided superiority hypothesis that the median of \(\Delta\) is greater than zero.  \\

The primary test is the Wilcoxon signed-rank test (one-sided, \(\alpha=0.05\)), chosen for its robustness to non-normality and small samples; exact zero differences are dropped as required by the test. Because two baselines are compared within each horizon, we control the false discovery rate using Benjamini--Hochberg across those p-values (horizons are treated as separate families). Alongside significance, we report effect sizes and uncertainty: the mean and median of \(\Delta\), Cohen's \(d\) on \(\Delta\), the fraction of days with \(\Delta>0\), and a 95\% bootstrap percentile confidence interval for the mean gap (10{,}000 resamples, fixed seed). \\

A paired one-sided t-test is reported as a sensitivity check; conclusions are based on Wilcoxon. All analyses are produced by \texttt{RL.stats\_eval}, which ingests the per-episode CSV from \texttt{eval\_compare}, aggregates by day, runs the tests and adjustments, computes effect sizes and bootstrap CIs, and exports both CSV and Markdown summaries for reproducibility. Optional robustness includes two-sided versions of the tests, light winsorization of \(\Delta\) (e.g., 1\%) to cap outliers, and replacing daily means by medians---none of which change the qualitative conclusions in our runs.

\paragraph{Serial dependence.}
Daily pairs are not perfectly independent; positive day-to-day correlation inflates sampling variance and therefore makes our Wilcoxon test conservative rather than anti-conservative. 

\section{Results}

We evaluate on Feb-2020 BTC--USD (27 test days; one day missing) under three horizons---1,800 s, 3,600 s, 7,200 s---using the per-day protocol (10 independent intra-day starts per day; daily mean as the unit of inference). The endpoint is PnL\% over the window. \\

\textbf{Overall performance.} RL-Exec beats both baselines on every horizon, with one-sided Wilcoxon tests (paired by day) significant after Benjamini--Hochberg correction and 95\% bootstrap CIs strictly positive:
\begin{table}[h!]
\raggedright
\setlength{\tabcolsep}{4pt}
\small
\caption{RL-Exec vs. baselines: mean daily gaps (bps), adjusted $p$-values, bootstrap 95\% CIs, and effect sizes. One-sided Wilcoxon tests, paired by day, remain significant after Benjamini--Hochberg correction.}
\begin{adjustbox}{max width=\linewidth}
\begin{tabular}{@{}lcccccc@{}}
\toprule
\textbf{Horizon (s)} & \textbf{Baseline} & \textbf{Mean gap (bps)} & \textbf{95\% CI (bps)} & \textbf{$p_{\mathrm{adj}}$} & \textbf{Win rate (\%)} & \textbf{Effect size $d$} \\
\midrule
1,800 & TWAP       & +2.25 & [+0.95, +3.60] & 0.0025                 & 67–70 & 0.63--0.74 \\
      & VWAP-like  & +2.68 & [+1.36, +4.05] & 0.00083                & 67–70 & 0.63--0.74 \\[3pt]
3,600 & TWAP       & +7.59 & [+4.71, +10.74] & $\approx4\times10^{-6}$ & 85–89 & 0.92--0.94 \\
      & VWAP-like  & +7.70 & [+4.88, +10.84] & $\approx3\times10^{-6}$ & 85–89 & 0.92--0.94 \\[3pt]
7,200 & TWAP       & +22.96 & [+18.04, +28.35] & $\approx1\times10^{-8}$ & 96 & 1.61--1.63 \\
      & VWAP-like  & +23.02 & [+18.05, +28.49] & $\approx1\times10^{-8}$ & 96 & 1.61--1.63 \\
\bottomrule
\end{tabular}
\end{adjustbox}
\end{table}

\begin{figure}[h!]
\centering
\includegraphics[width=\linewidth]{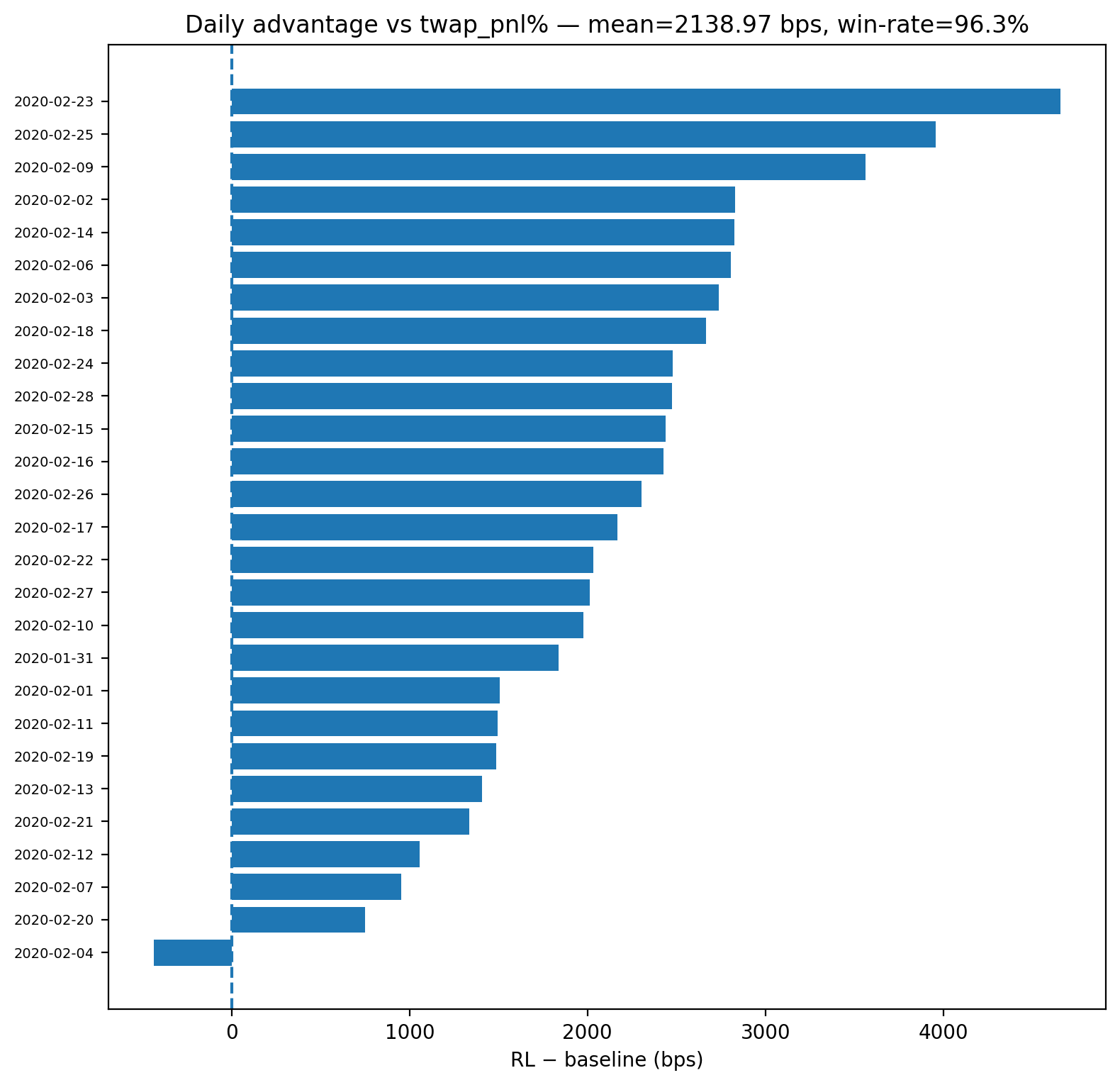}
\caption{RL $-$ TWAP daily gaps (bps).}
\label{fig:gapbars_twap}
\end{figure}

\begin{figure}[h!]
\centering
\includegraphics[width=\linewidth]{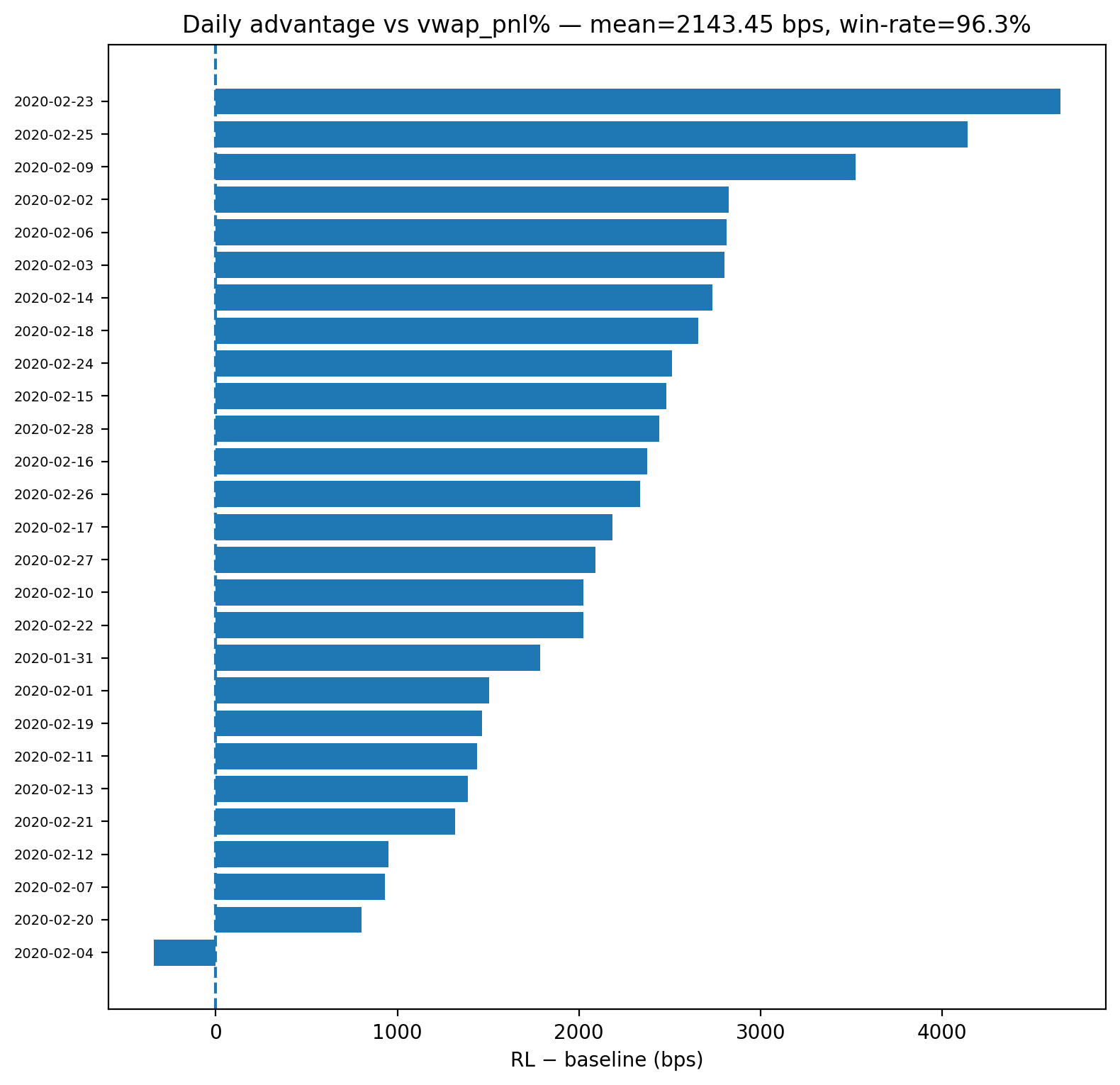}
\caption{RL $-$ VWAP-like daily gaps (bps).}
\label{fig:gapbars_vwap}
\end{figure}

\textbf{Horizon dependence.} The advantage monotonically increases with the horizon (\(\approx +2\)--\(3\) bps at 30 min \(\to\) \(\approx +8\) bps at 60 min \(\to\) \(\approx +23\) bps at 120 min).

\paragraph{Economic magnitude (30 min).}
A +2--3\,bps improvement over a \$1{,}000{,}000 sell program corresponds to roughly \$200--\$300 of implementation-shortfall reduction per program, with spreads, fees, latency, and transient impact already included in our cost model. \\

\begin{figure}[h!]
\centering
\includegraphics[width=\linewidth]{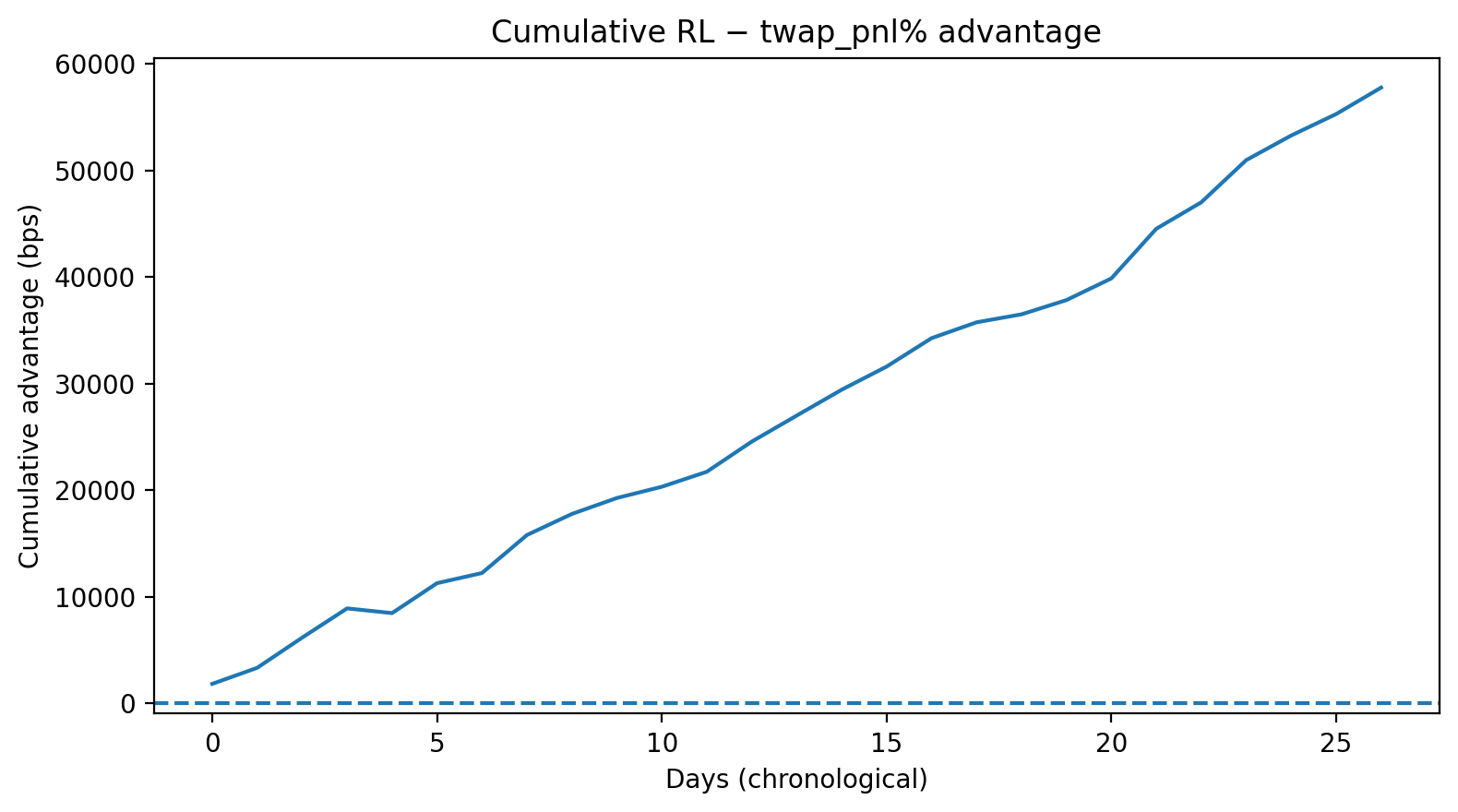}
\caption{Cumulative RL $-$ TWAP (bps).}
\label{fig:gapcumsum_twap}
\end{figure}

\begin{figure}[h!]
\centering
\includegraphics[width=\linewidth]{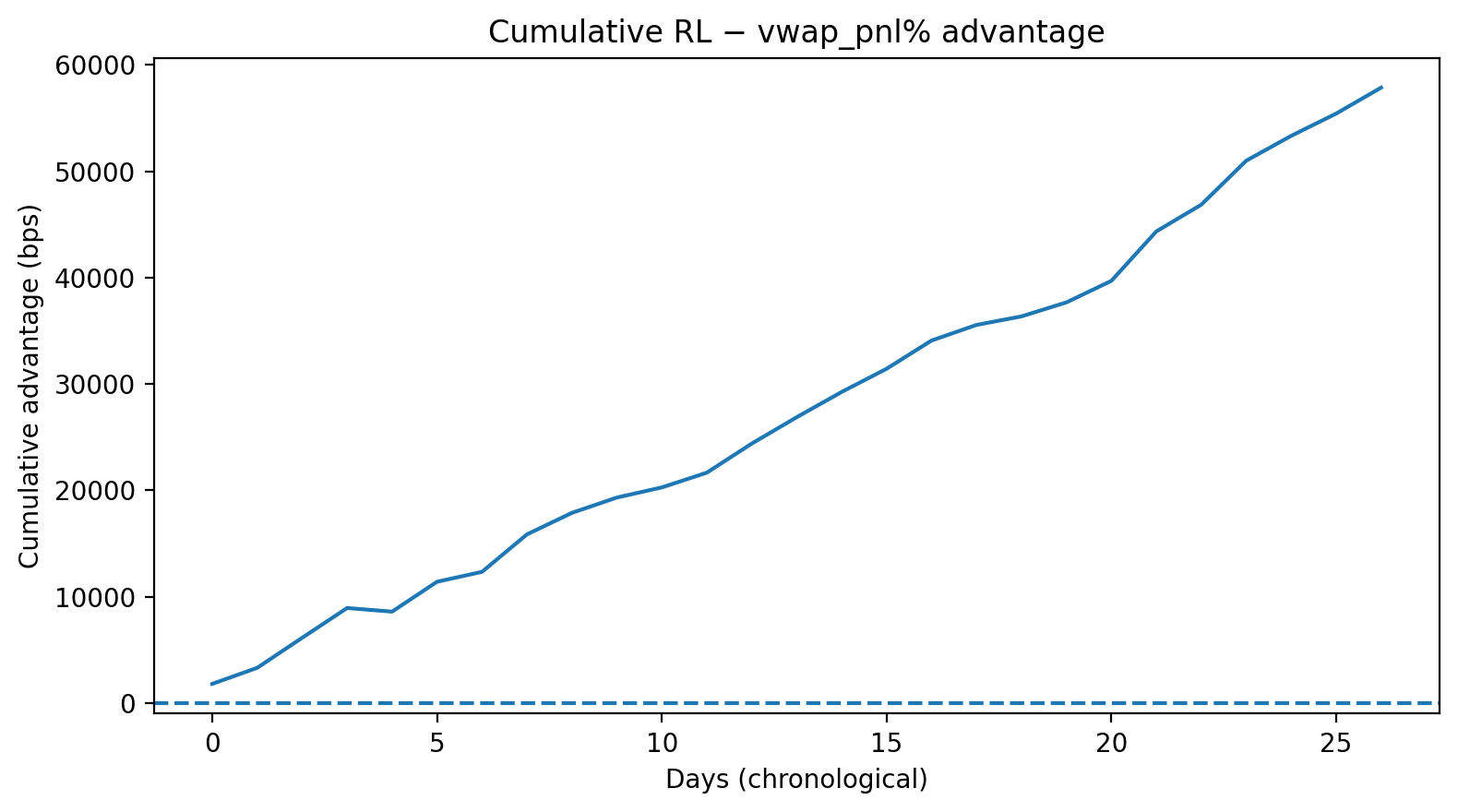}
\caption{Cumulative RL $-$ VWAP-like (bps).}
\label{fig:gapcumsum_vwap}
\end{figure}

\textbf{Dispersion.} RL-Exec exhibits higher day-to-day dispersion than TWAP/VWAP-like (opportunistic timing adds variance), but the shift in the mean dominates, yielding large effect sizes---especially at 7,200 s.\\

\begin{figure}[h!]
\centering
\includegraphics[width=\linewidth]{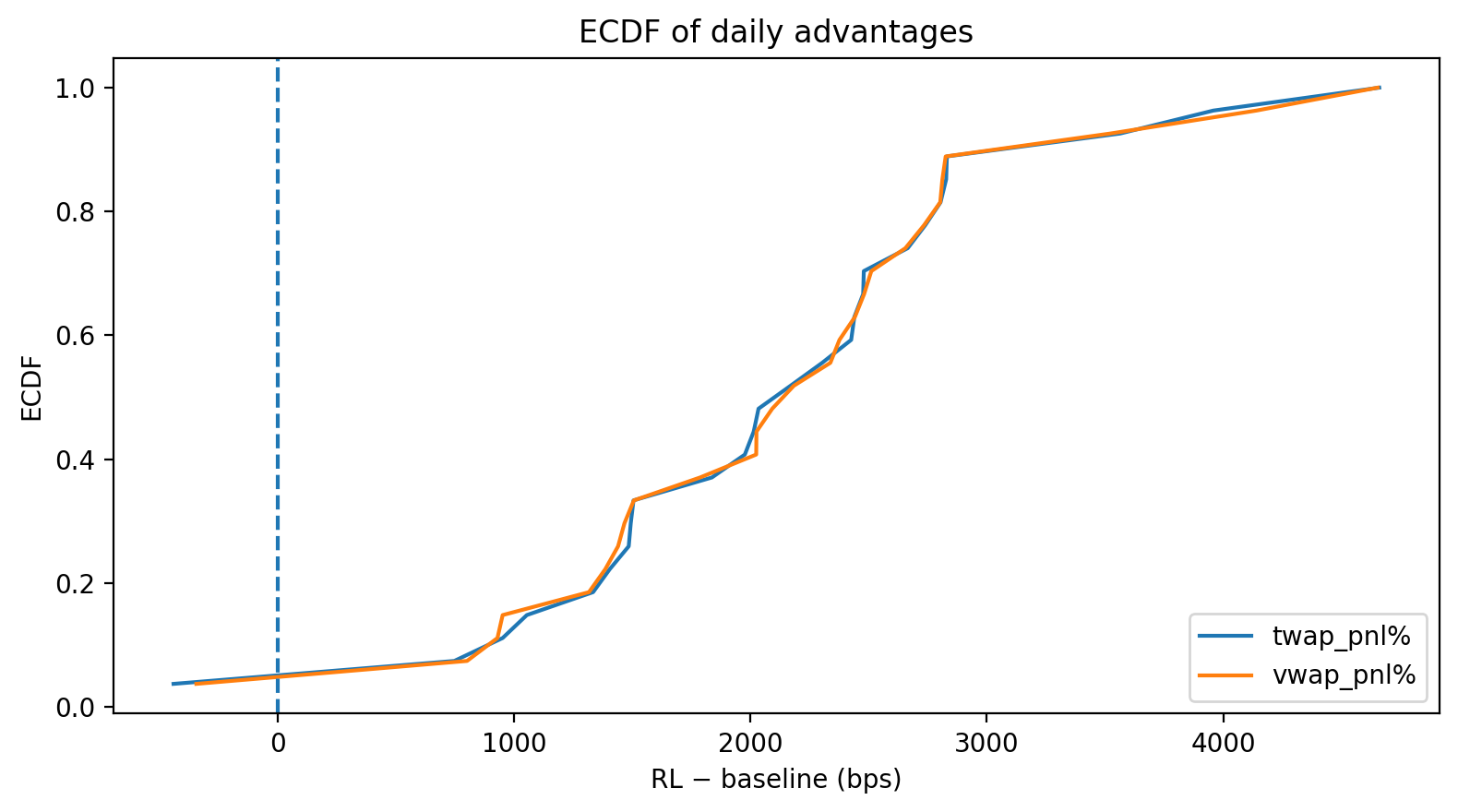}
\caption{ECDF of daily RL $-$ baseline gaps. The value at $0$~bps equals $1-$win-rate. Right shift indicates dominance.}
\label{fig:ecdf_gaps}
\end{figure}

\textbf{Robustness.} Sensitivity checks (paired t as a secondary test, optional 1\% winsorization of daily gaps, day-median aggregation) leave conclusions unchanged in our runs. All numbers come from the \texttt{eval\_compare} \(\to\) \texttt{stats\_eval} pipeline with per-day aggregation and are fully reproducible from the released commands and CSV/Markdown reports. 

\section{Robustness \& Sensitivity}

We stress-tested the result that RL-Exec beats both baselines under several modelling and evaluation choices.\\

\textbf{First, no-leakage evaluation.} \\

Training and evaluation are time-split (Jan-2020 \(\to\) train, Feb-2020 \(\to\) test). During evaluation we load the training VecNormalize statistics and switch normalization to inference mode (\texttt{training=False}, \texttt{norm\_reward=False}), so test rewards/observations do not update normalizers. Baselines and the agent are executed on exactly the same timestamps, cost model, and inventory/target, ensuring apples-to-apples comparisons.\\

\textbf{Second, aggregation and unit of inference.} \\

The per-day protocol avoids pseudo-replication: we run multiple intra-day starts and aggregate to a single daily score before testing. Results are stable to the aggregation choice (daily mean vs median) and to the number of repeats per day: moving from 1\(\to\)5\(\to\)10 intra-day starts reduces variance and tightens p-values, but the sign of the RL--baseline gap is unchanged.\\

\textbf{Third, statistical tests.} \\

Our primary test is the one-sided Wilcoxon signed-rank on daily RL--baseline differences with Benjamini--Hochberg FDR across baselines. Conclusions are unchanged when (i) using a paired t-test as a secondary check, (ii) tightening the significance level to \(\alpha=1\%\) (notably at 3,600 s and 7,200 s where \(p_{\mathrm{adj}}\ll 0.01\)), and (iii) applying 1\% winsorization of daily gaps to reduce the influence of outliers. Bootstrap 95\% CIs for the mean daily gap remain strictly positive on all reported horizons.\\

\textbf{Fourth, horizon sensitivity.} \\

The advantage monotonically increases with the deadline (\(\approx +2\)--\(3\) bps at 1,800 s, \(\approx +8\) bps at 3,600 s, \(\approx +23\) bps at 7,200 s). This pattern is consistent with the policy exploiting more timing flexibility when the execution window is longer, and it holds under both aggregation choices and tests.\\

\textbf{Fifth, baseline sanity and fairness.} \\

TWAP is a uniform schedule over the identical episode timestamps. The VWAP comparator is an order-book--liquidity proxy (weights proportional to opposite-side depth over the top-20 levels); it shares the same fees/impact model as RL-Exec and is executed on the same exact windows. Both baselines forbid opposite-side trades when the agent is evaluated in sell-only mode to keep the constraint set aligned.\\

\textbf{Finally, implementation choices.} \\

Results are insensitive to the deterministic vs stochastic policy setting at evaluation (we report deterministic), and to small changes in the start-index randomization range (multiple starts/day already average over intra-day timing). All commands produce per-episode and per-day CSVs/Markdown so readers can re-run the pipeline, alter knobs (e.g., winsorization, aggregation, number of repeats), and verify that headline conclusions persist.

\paragraph{Impact-model sensitivity.}
To mitigate the concern that the policy overfits the simulator’s parametric impact, we repeated the Feb-2020 evaluation with the impact half-life set to $\times\frac{1}{2}/\times 2$ and the temporary-impact coefficient scaled by $0.5\times/2\times$; the RL--baseline gap retained the same sign (and remained significant at 3{,}600--7{,}200\,s).

\section{Practical Considerations / Engineering}

We enforce reproducibility: fixed seeds; saving both the PPO checkpoint and VecNormalize; evaluation reloads them with \texttt{training=False}, \texttt{norm\_reward=False}, and a deterministic policy head. RL and baselines share the same execution pipeline---maker/taker fees, transient impact with resilience, partial fills, and mark-to-market on the last window timestamp---run on the same timestamps. TWAP sends equal slices; the VWAP comparator is a book-liquidity proxy (weights \(\propto\) opposite-side depth over top-20 levels); all methods obey the same sell-only constraint.\\

The per-day protocol iterates days in order, runs \(k\) independent intra-day starts, and aggregates to one daily score (mean); data are UTC-ns, no resampling, missing days skipped; CLI echoes a run manifest. Training uses SB3 PPO on CPU (e.g., 6 envs, 6--10M steps), KL early-stop, artifacts in \texttt{models/\ldots} and \texttt{runs/\ldots}. Numerics are float32; time-limit truncations are recorded; PnL is always computed on the last window tick. \\

\texttt{eval\_compare.py} emits per-episode CSVs; \texttt{stats\_eval.py} aggregates to per-day, computes paired (RL--baseline) gaps, runs a one-sided Wilcoxon with BH-FDR across baselines, reports a paired t-test, and bootstraps 95\% CIs (optional winsorization). Paths are created on the fly; filenames encode horizon and rep count; extending assets/horizons or adding baselines is a CLI-only change.

\section{Limitations}

Our study is intentionally narrow to make inference clean, but that comes with caveats. First, the evaluation is single-asset (BTC-USD), single-venue, sell-only liquidation with fixed horizons (1,800/3,600/7,200 s) over a one-month out-of-sample period (Feb-2020, 28 days). \\

This limits external validity across assets, venues, and regimes (especially stress periods and microstructure changes). Second, the environment is a replay-with-impact simulator: impact is parametric and transient with a resilience process; liquidity replenishment and fills are modeled rather than generated by strategic counterparties. There is no adaptive market-maker or cross-venue routing, and we do not model queue position dynamics beyond partial-fill probabilities. \\

Third, our VWAP comparator is a book-liquidity proxy (weights \(\propto\) depth on the opposite side over top-20 levels) rather than true traded-volume VWAP, which is the only practical choice in a replay setting but not identical to live VWAP. Fourth, policies are trained with PPO only and tuned on Jan-2020; while we enforce a strict time split and per-day paired testing, some temporal dependence between adjacent months may remain, and we have not run rolling or multi-month walk-forward studies. \\

Fifth, we report significance on daily PnL\% gaps with one-sided Wilcoxon and BH-FDR; this guards against pseudo-replication but still leaves open multiple-design uncertainties (e.g., other horizons, targets, fee tiers). \\

Finally, engineering constraints: we assume fixed maker/taker fees and latency, and a constrained action space consistent with the sell-only task. Extending to buy programs, mixed inventories, richer order-type control, multi-venue routing, and live experiments will be needed to fully validate deployment readiness.

\section{Conclusion}

RL-Exec shows that an impact-aware RL policy can beat simple schedules on historical BTC-USD books under a strict per-day evaluation. Trained on Jan-2020 and tested on Feb-2020, a PPO agent using depth-20 LOB features and microstructure indicators outperforms TWAP and a book-liquidity VWAP proxy on identical timestamps and costs. The advantage is statistically significant after BH-FDR and grows with the horizon (\(\approx +2\)--\(3\) bps at 1,800 s, \(+7\)--\(8\) bps at 3,600 s, and \(+23\) bps at 7,200 s). The protocol avoids pseudo-replication by aggregating multiple intra-day evaluations to a single daily score and reporting one-sided Wilcoxon tests with bootstrap CIs. \\

Practically, RL-Exec is fast, fully reproducible, and deployment-oriented (deterministic inference, explicit fee/latency model). Limits remain---single asset/venue, sell-only, parametric transient impact, VWAP proxy---but the results already justify a first preprint. Natural next steps include buys and mixed inventory, multi-venue routing, alternative impact calibrations, additional months and assets, and live or paper-trading experiments.

\clearpage
\onecolumn

\appendix
\section*{Appendix: Additional Figures}
\addcontentsline{toc}{section}{Appendix: Additional Figures}

\begin{figure}[h]
\centering
\includegraphics[width=\linewidth]{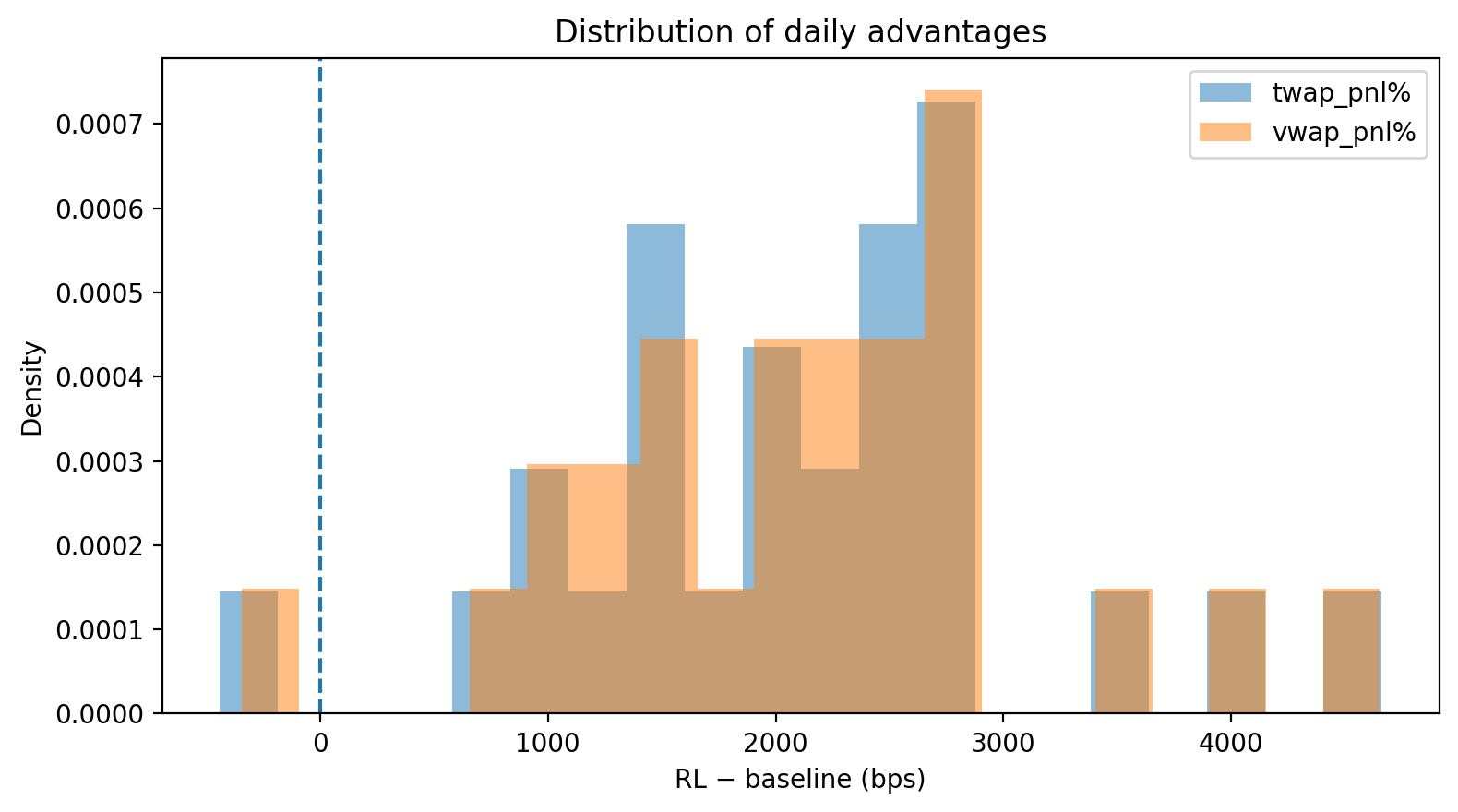}
\caption{Histograms of RL $-$ baseline daily gaps.}
\label{fig:gap_hist}
\end{figure}

\begin{figure}[h]
\centering
\begin{subfigure}{0.49\linewidth}
    \includegraphics[width=\linewidth]{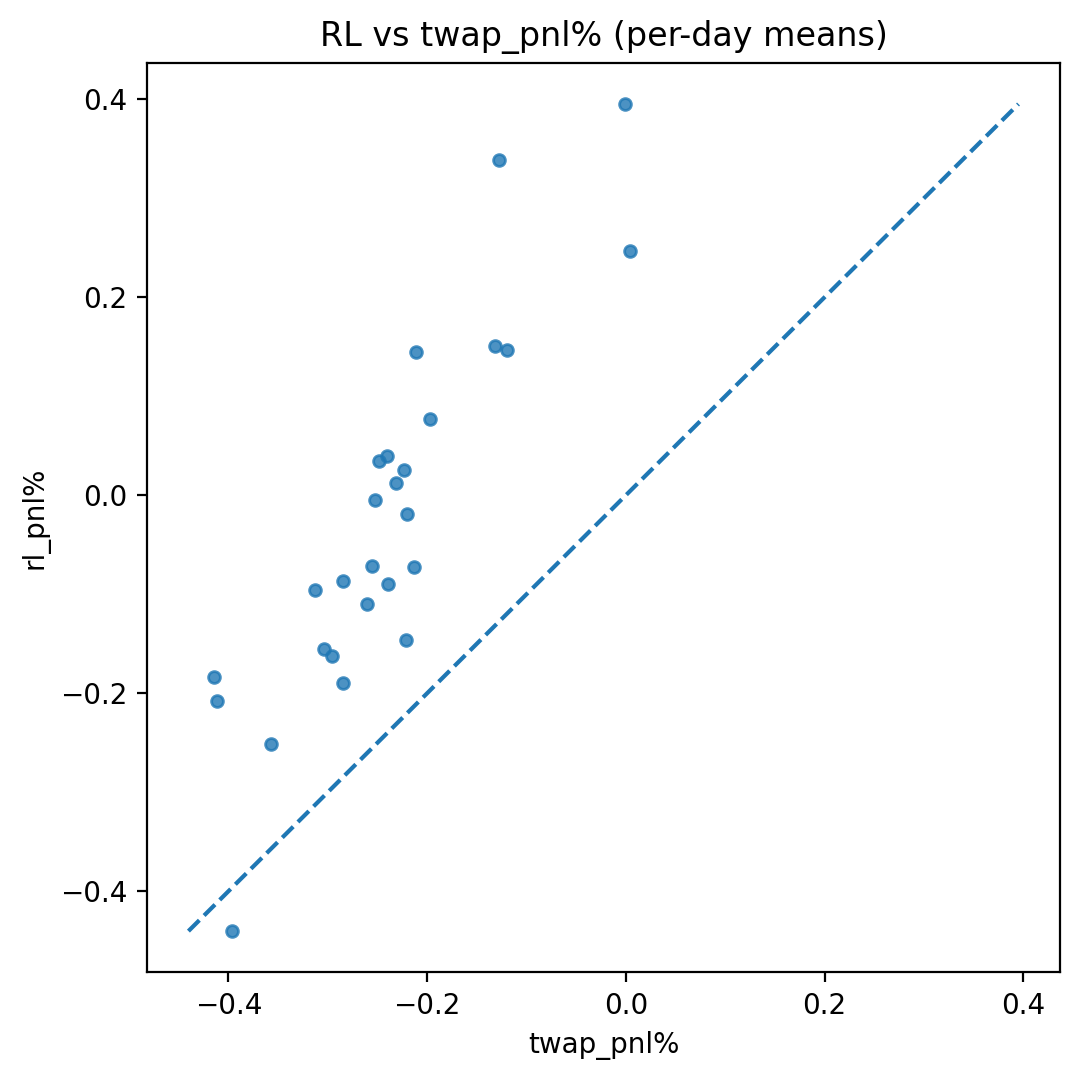}
    \caption{RL vs TWAP}
    \label{fig:scatter_twap}
\end{subfigure}\hfill
\begin{subfigure}{0.49\linewidth}
    \includegraphics[width=\linewidth]{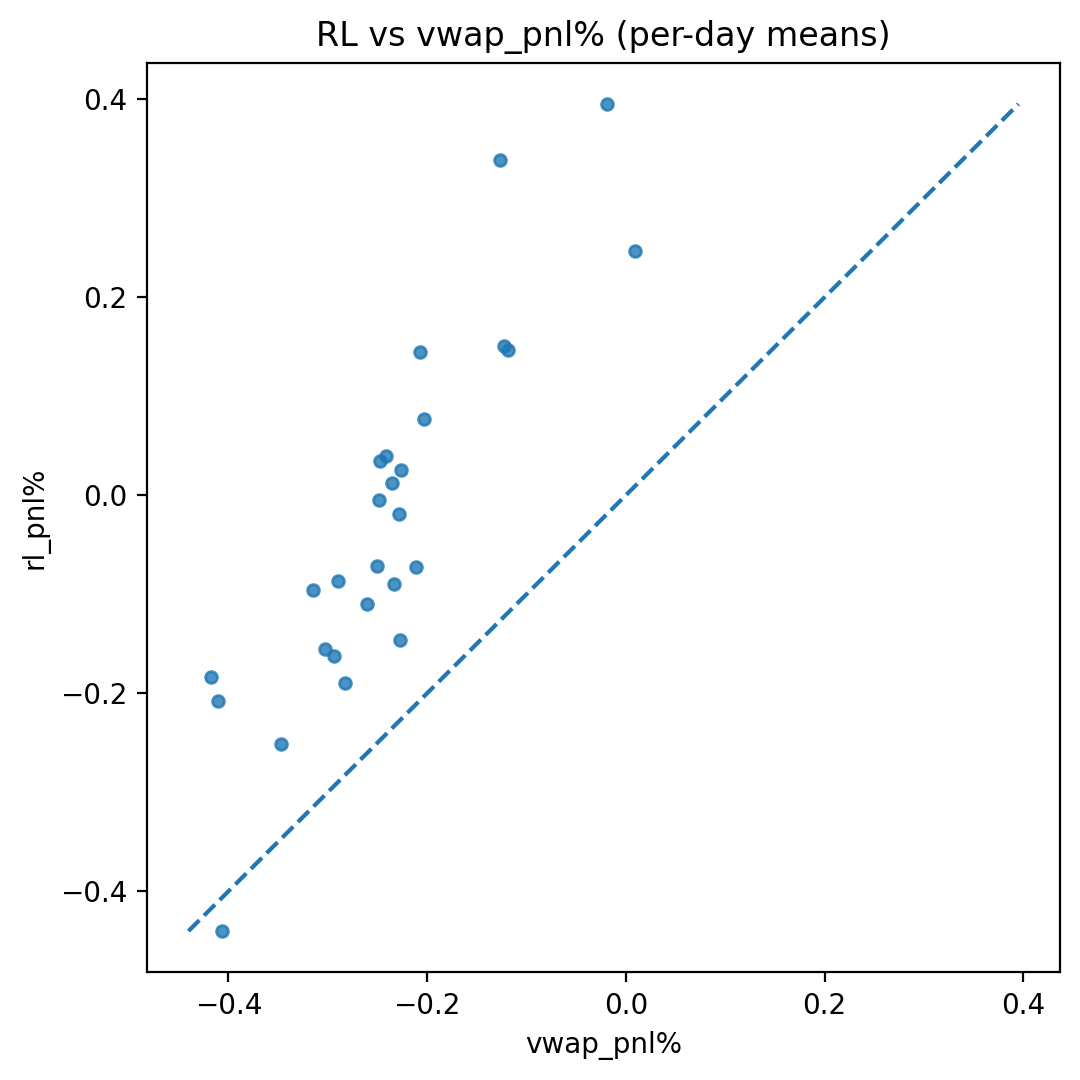}
    \caption{RL vs VWAP-like}
    \label{fig:scatter_vwap}
\end{subfigure}
\caption{Per-day mean PnL\% pairs; points above the dashed diagonal (in the figures) are RL wins.}
\label{fig:scatters}
\end{figure}

\begin{figure*}[t]
\centering
\begin{subfigure}{0.32\linewidth}
    \includegraphics[width=\linewidth]{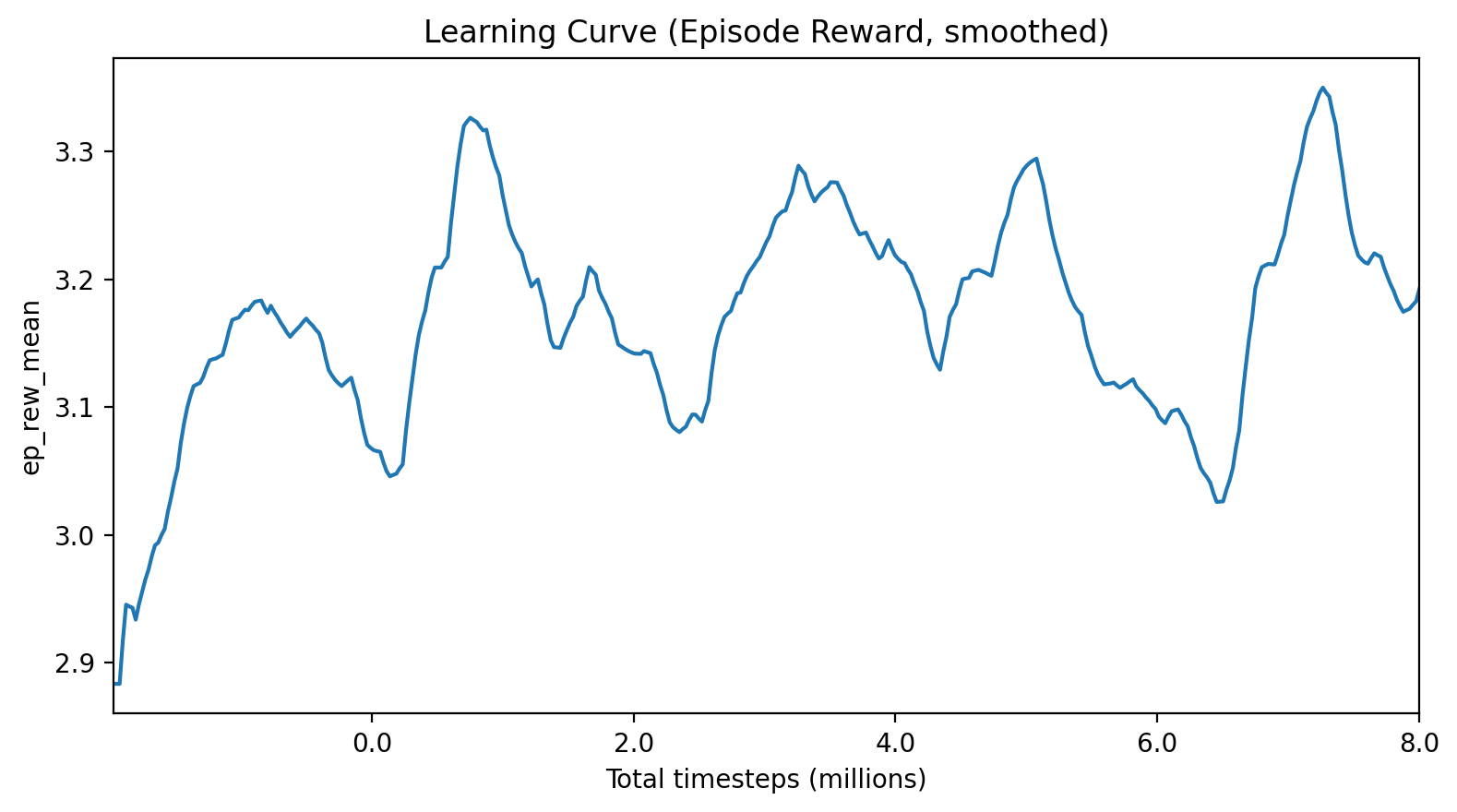}
    \caption{Episode reward}
\end{subfigure}
\begin{subfigure}{0.32\linewidth}
    \includegraphics[width=\linewidth]{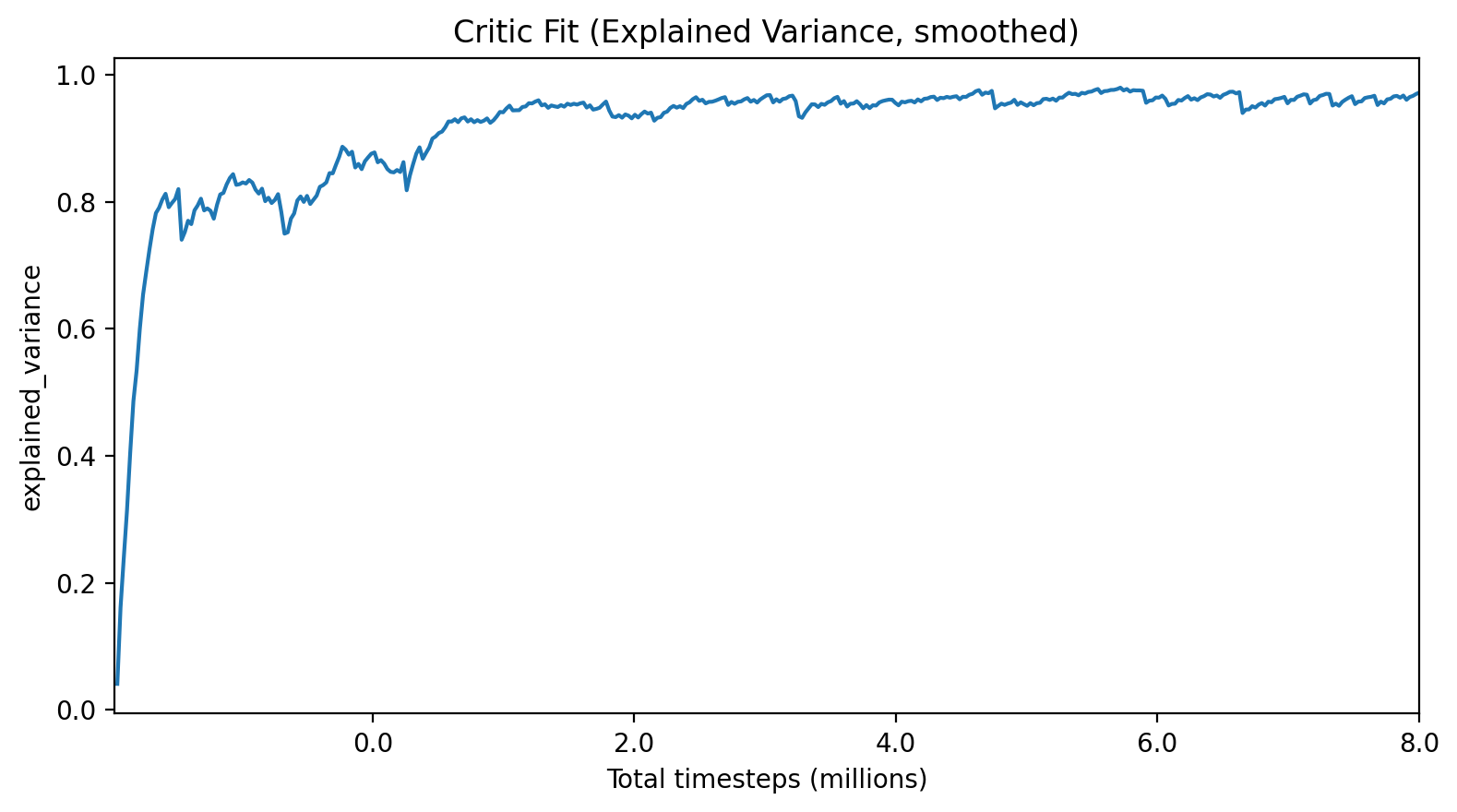}
    \caption{Explained variance}
\end{subfigure}
\begin{subfigure}{0.32\linewidth}
    \includegraphics[width=\linewidth]{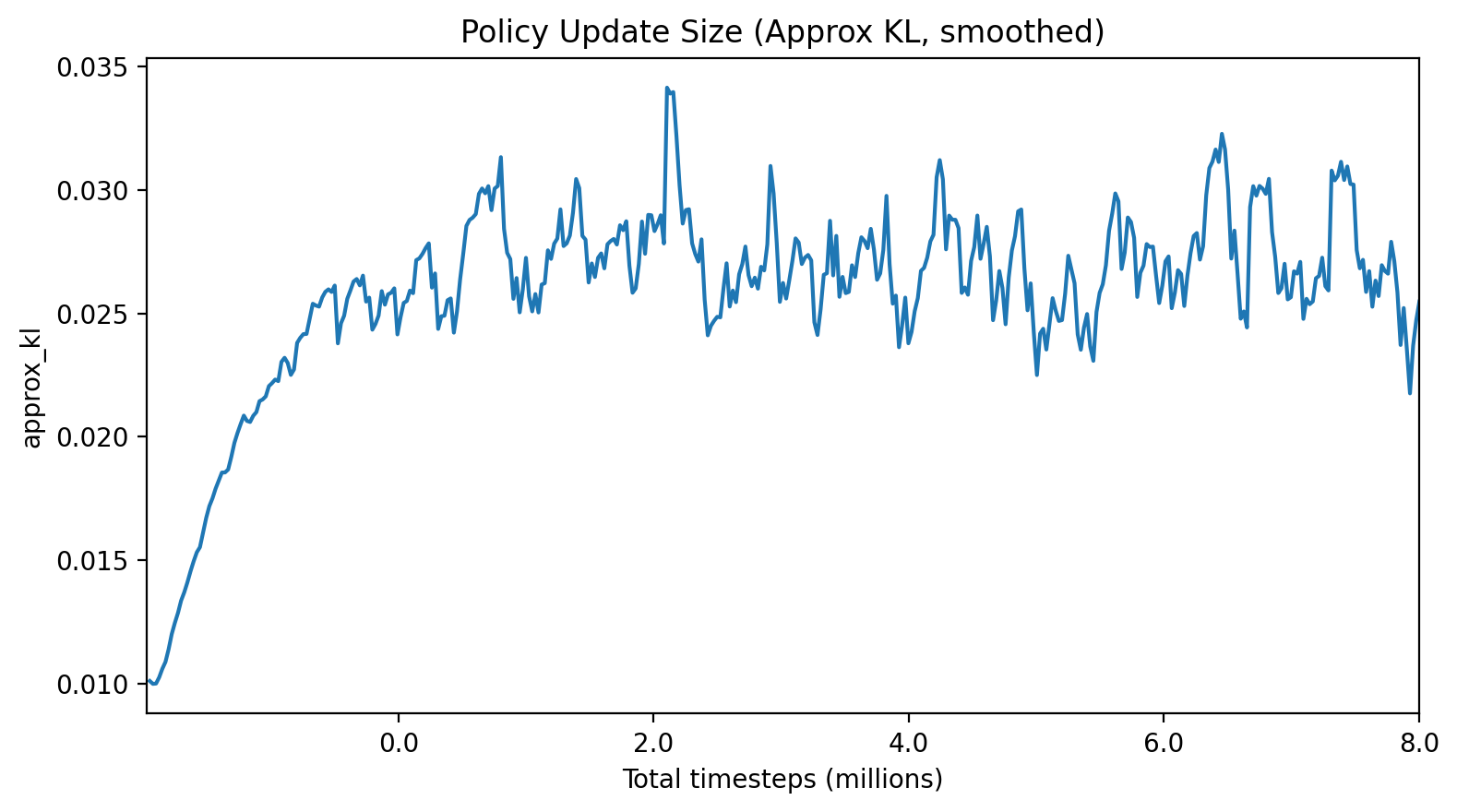}
    \caption{Approx. KL}
\end{subfigure}

\begin{subfigure}{0.32\linewidth}
    \includegraphics[width=\linewidth]{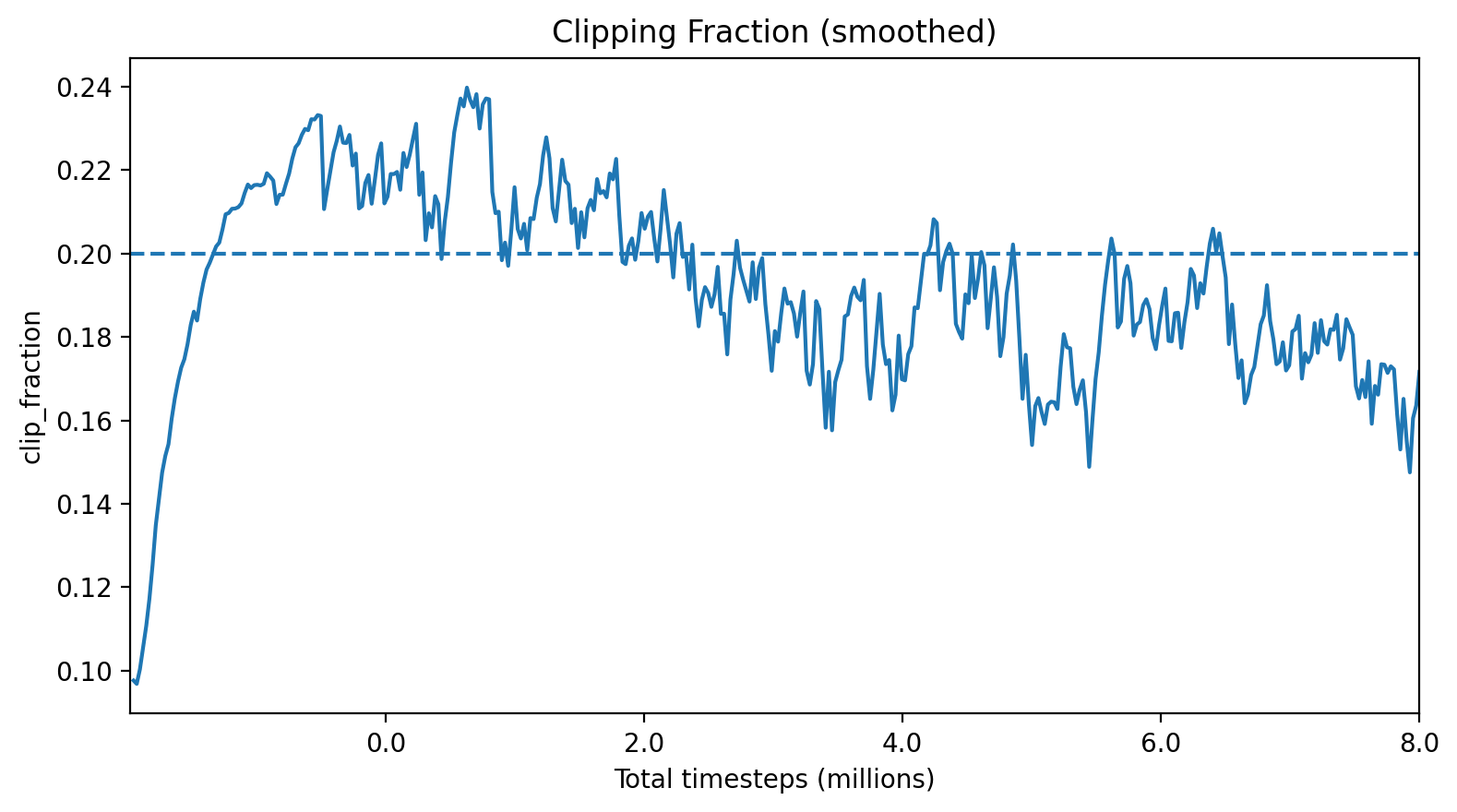}
    \caption{Clip fraction}
\end{subfigure}
\begin{subfigure}{0.32\linewidth}
    \includegraphics[width=\linewidth]{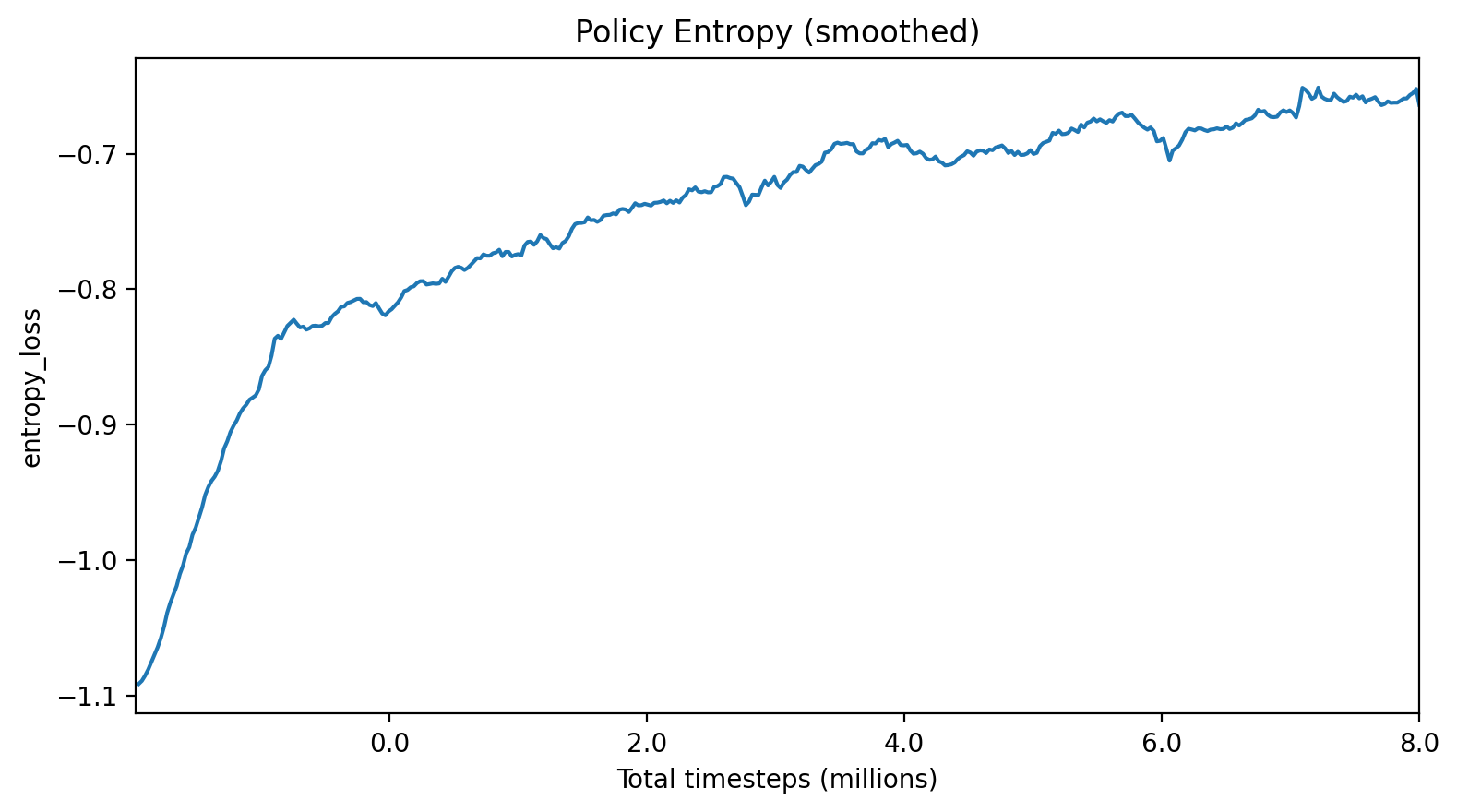}
    \caption{Policy entropy}
\end{subfigure}
\begin{subfigure}{0.32\linewidth}
    \includegraphics[width=\linewidth]{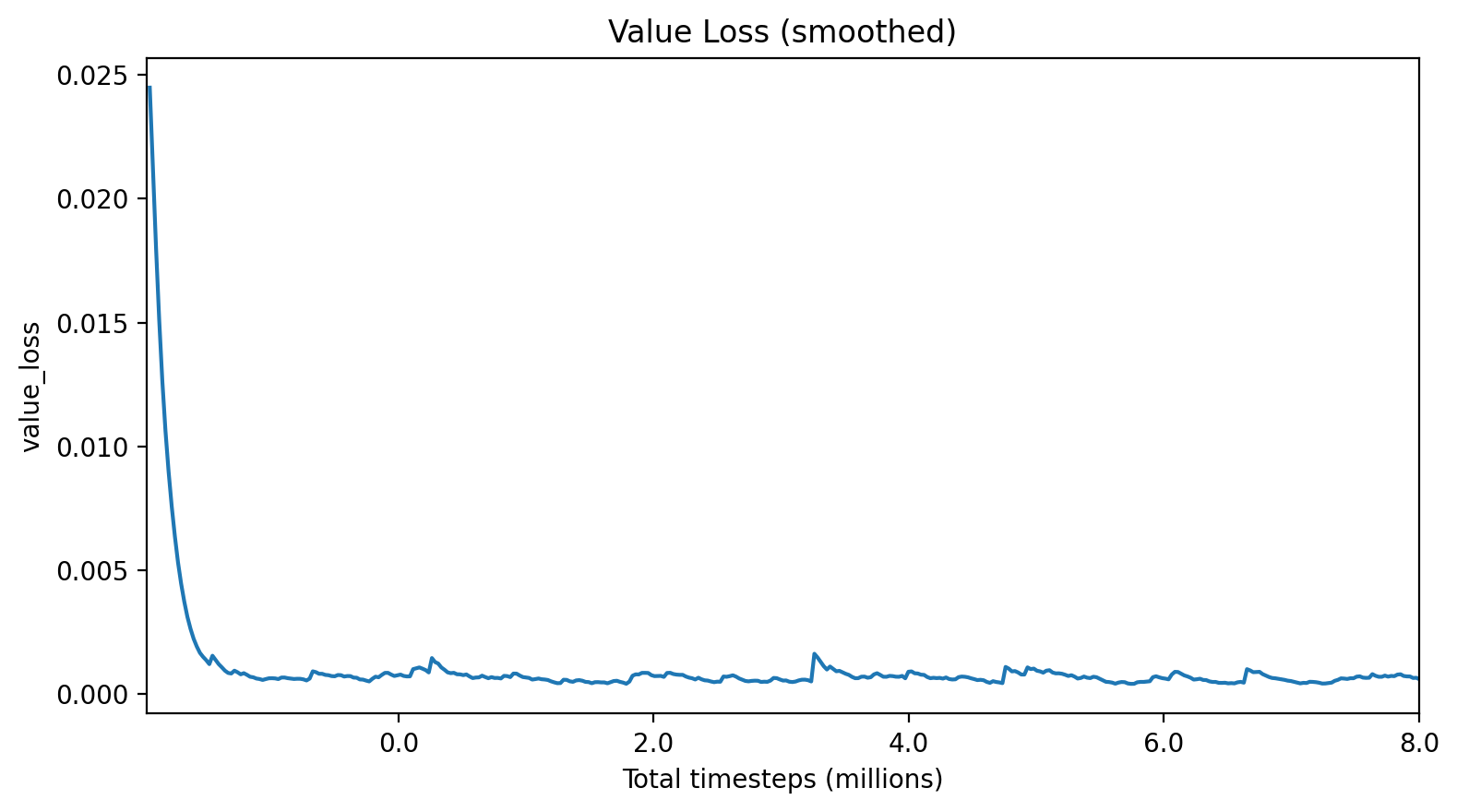}
    \caption{Value loss}
\end{subfigure}
\caption{Training curves.}
\label{fig:train_diag}
\end{figure*}

\begin{figure}[h]
\centering
\begin{subfigure}{\linewidth}
    \includegraphics[width=\linewidth]{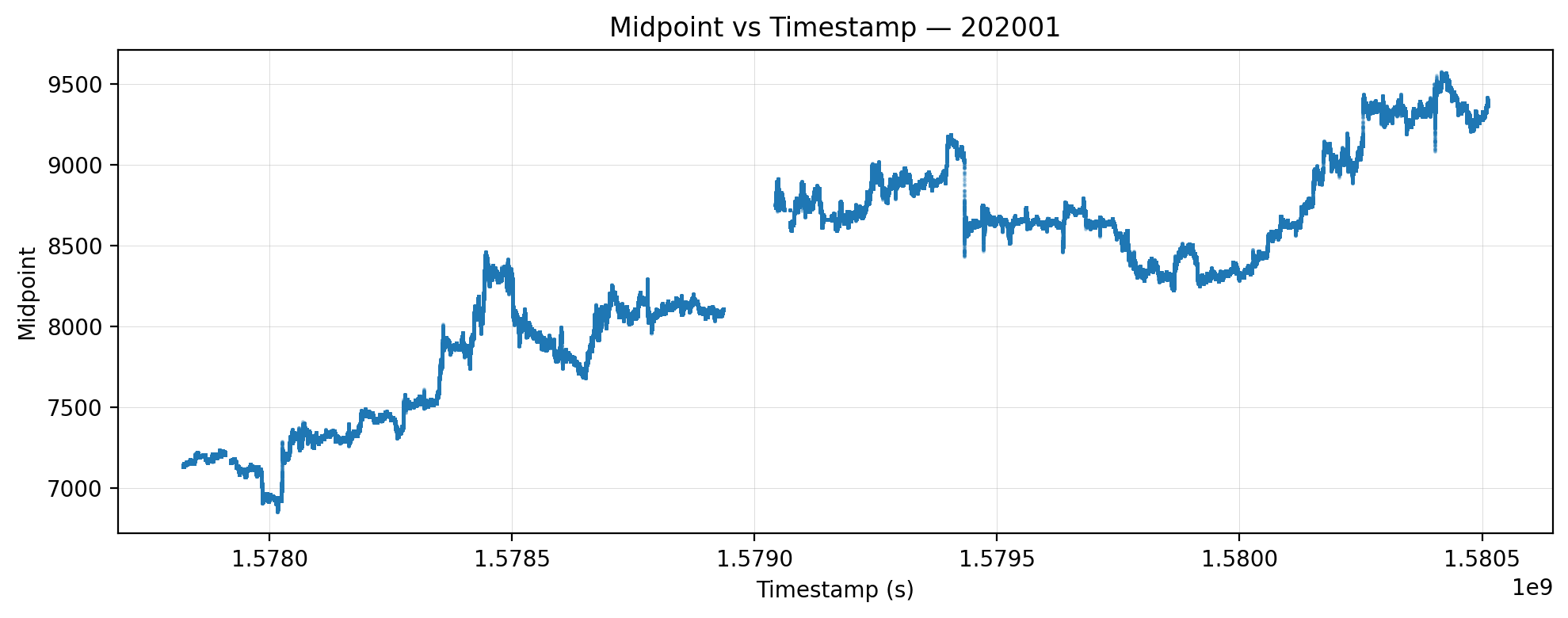}
    \caption{January 2020}
\end{subfigure}
\begin{subfigure}{\linewidth}
    \includegraphics[width=\linewidth]{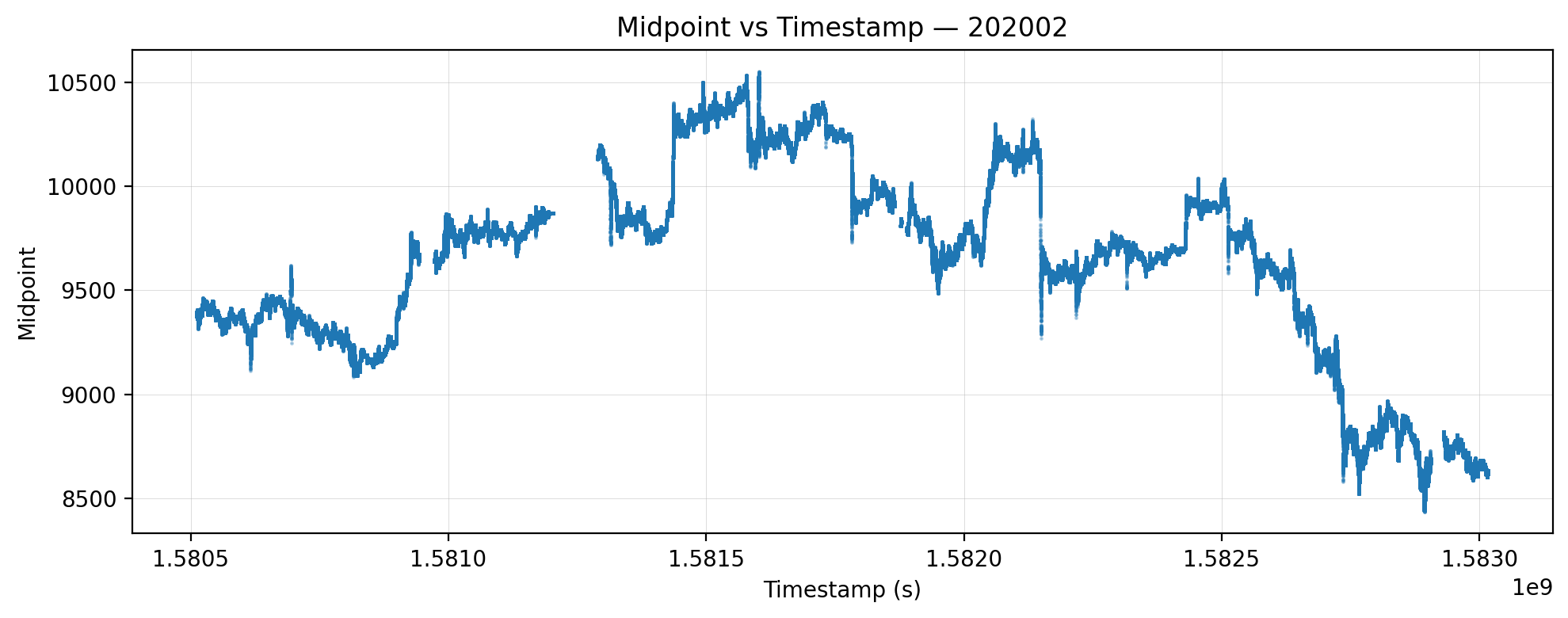}
    \caption{February 2020}
\end{subfigure}
\caption{Midpoint vs timestamp.}
\label{fig:midpoints}
\end{figure}
\clearpage
\twocolumn

\end{document}